# Determining the unsaturated hydraulic conductivity of a compacted sand-bentonite mixture under constant volume and free-swell conditions

**Y.J. Cui, A.M. Tang, C. Loiseau, P. Delage**

Université Paris Est, Ecole des ponts, I. Navier

**Corresponding author**

**Prof. Yu-Jun Cui**

ENPC/CERMES
6 et 8 avenue Blaise Pascal,
Cité Descartes, Champs-sur-Marne,
77455 MARNE-LA-VALLEE CEDEX 2,
France.
Tel: 33 1 64 15 35 50
Fax: 33 1 64 15 35 62
E-mail: cui@cermes.enpc.fr

**Abstract**

Highly compacted sand-bentonite mixtures are often considered as possible engineered barriers in deep high-level radioactive waste disposals. In-situ, the saturation of these barriers from their initially unsaturated state is a complex hydro-mechanical coupled process in which temperature effects also play a role. The key parameter of this process is the unsaturated hydraulic conductivity of the barrier. In this paper, isothermal infiltration experiments were conducted to determine the unsaturated hydraulic conductivity according to the instantaneous profile method. To do so, total suction changes were monitored at different locations along the soil specimen by using resistivity relative humidity probes. Three constant volume infiltration tests were conducted showing, unexpectedly, a decrease of the hydraulic conductivity during infiltration. One test performed under free-swell conditions showed the opposite and standard trend. These observations were interpreted in terms of microstructure changes during wetting, both under constant volume and free swell conditions.

## *Introduction*

In radioactive waste disposal at great depth, compacted expansive soils are sometimes considered as possible engineered barrier to be placed between the radioactive waste and the host rock. When hydrated by the pore water infiltrated from the host rock, engineered barriers cannot swell, resulting in the development of swelling pressure, microstructure changes and related changes in hydraulic properties. A proper understanding of this particular situation is necessary in the description of water transfers in a real storage condition.

The saturated hydraulic conductivity ($k_{sat}$) of bentonite-based buffer materials is often measured by hydrating the specimen under constant volume condition with a constant inlet water pressure and by monitoring the water inlet and/or outlet flow. $k_{sat}$ is calculated using Darcy's law. It has been observed that $k_{sat}$ decreased with increased dry density (Dixon et al. 1999), increased bentonite content (Komine 2004; Kenney et al. 1992) or decreased temperature (in the range of 20 – 80 °C, after Cho et al. 1999, Villar and Lloret 2004). In addition, Pusch (1982), Haug and Wong (1992) and Loiseau et al. (2002) observed that water infiltration at constant volume modified the soil microstructure, giving rise to $k_{sat}$ decrease.

The unsaturated hydraulic conductivity ($k_{unsat}$) can be measured in the laboratory by various methods among which unsteady methods such as the instantaneous profile method (Daniel 1982) are the most suitable for clayey soils (Benson and Gribb 1997). For bentonite-based buffer material, $k_{unsat}$ is often determined from infiltration tests in columns. In some cases, compacted soil cylindrical specimens are wetted from one end and the water content profile is determined at a given time by cutting the specimen into slices (Börgesson et al. 2001; Kröhn 2003*a*). Lemaire et al. (2004) performed an infiltration test in an oedometer cell and monitored the change in water content and dry density with time at different locations by using dual-energy γ-ray measurement. Various methods have been applied to describe the water transfer obtained from infiltration tests: Among others, Lemaire et al. (2004) solved the common diffusion equation by using Boltzmann variable whereas Kröhn (2003*b*) considered both an advection model and a vapour diffusion model to fit his experimental data.

This paper presents the experimental results obtained on compacted sand/bentonite mixture specimens by performing three infiltration tests under constant volume conditions and one infiltration test under free-swell conditions. The instantaneous profile method was used to determine the unsaturated hydraulic conductivity. The results from the two conditions were compared to analyze the effect of microstructure changes on the hydraulic conductivity during wetting.

## Material and methods

The soil tested is a mixture of Kunigel-V1 bentonite and Hostun sand with a respective proportion of 7/3 in dry weight. According to Komine (2004), the Kunigel-V1 clay consists of 48% montmorillonite, resulting in high plasticity data (liquid limit $w_L$ = 474%, plasticity index $Ip$ = 447). The geotechnical properties of Kunigel-V1 bentonite (after Komine 2004) are presented in Table 1. The particle size distributions of Hostun sand used in the present work is given in Figure 1. The solid unit weight of the sand/bentonite mixture is 2.67 Mg/m$^3$. The mixture used in this work is similar to that used in the full-scale "Tunnel Sealing Experiment" (TSX test) conducted in Canada (Martino et al. 2007; Dixon et al. 2007).

Samples were prepared by compacting a mixture made up of clay powder and sand grains with an initial water content of 4.2%. Prior to compaction, the mixture was put in three airtight chambers at three different relative humidities ($RH$) controlled by using saturated saline solutions. At equilibrium, final water contents were respectively equal to 6.5±0.1% ($Mg(NO_3)_2$ solution, $RH$ = 55%, suction s = 82 MPa); 8.0±0.3% ($NaNO_2$ solution, $RH$ = 66%, suction s = 57 MPa); and 10.0±0.3% ($ZnSO_4$ solution, $RH$ = 90%, suction s = 12.6 MPa). The mixture was then compacted statically in a metallic cylinder of 50 mm diameter at a dry unit mass $\rho_d$ = 2.0 Mg/m$^3$. Loiseau et al. (2002) determined the water retention curves (suction s versus gravimetric water content w) of the mixture compacted at $w$ = 7.7% and $\rho_d$ = 2.0 Mg/m$^3$. The results obtained along the wetting path under free-swell and no volume changes conditions are presented in Figure 2.

In order to determine the unsaturated hydraulic conductivity of the compacted sand/bentonite mixture by using the instantaneous profile method, infiltration tests were performed under two conditions: constant-volume and free-swell. The device used for the infiltration test at constant-volume conditions is presented in Figure 3. The soil specimen (50 mm in diameter, 250 mm high) was directly compacted in the metallic cylinder (50 mm in inner diameter, 80 mm in outer diameter). Four resistive $RH$ sensors (Elcowa) were installed in small holes made into the sample through the four ports in the wall of the cylinder. The two ends of the cylinder were covered by two metallic discs of 40 mm thickness. The bottom of the cell was connected to a water source while the upper end was connected to an air source under atmospheric pressure. Porous stones were placed in both bottom and upper ends. One $RH$ probe was installed in the upper disk to monitor the $RH$ on the top of the soil specimen. The device enables $RH$ monitoring at five different distances from the wetted end ($h$ = 50, 100, 150, 200, and 250 mm).

The infiltration test with swelling allowed is schematically described in Figure 4 and a picture is presented in Figure 5. In this system, the cylindrical compacted soil specimen (50 mm in diameter, 100 mm high) was wrapped by a deformable neoprene membrane (0.3 mm thick) and placed horizontally on a bed made up of glass balls in order to reduce friction between the sample and its support during swelling. Four $RH$ probes were embedded into the sample at distances $h$ = 10, 40, 65, and 95 mm from the wetting end. The connecting wires went through the membrane and air-tightness was ensured by putting silicon glue between the membrane and the wire. Four displacement transducers (0.001 mm accuracy) were also used to monitor the local radial swelling of the soil specimen at distances $h$ = 9, 36, 71.5 and 96.5 mm from the wetted end. A displacement transducer was also installed on the top of soil specimen to monitor the total axial swelling. In order to better visualize the soil swelling, a grid (2x2 cm) was plotted on the membrane. The volumes of inlet water and expelled air were equally monitored using graduated fine tubes.

Three tests were performed at constant volume with three initial water contents ($w_i$ = 7.70% - Test T01, $w_i$ = 6.45% - Test T02, and $w_i$ = 9.95% - Test T03). The test performed with swelling allowed (Test T04) had $w_i$ = 8.20%.

## *Experimental results*

Figure 6 presents the *RH* changes with time of the three tests performed at constant volume in the column of Figure 3. The initial *RH* value of Test T01 ($w_i$ = 7.70 %) was 70±1% (suction equal to 49.5 ± 1.5 MPa). To reach this water content, the sand/bentonite mixture was previously hydrated in the vapour phase in an airtight chamber at *RH* = 66%. The *RH* of the soil increased from 66 to 70% during subsequent compaction. Once infiltration started, the *RH* value at $h$ = 50 mm increased rapidly and reached a zero suction state (*RH* $\cong$ 100%) after 2500 h. For other *RH* probes, the further the distance from the wetting face, the slower the rate of *RH* increase. For instance, *RH* at the upper end of the soil specimen ($h$ = 250 mm) increased only from 70 to 73% after 2500 h infiltration.

In Test T02 ($w_i$ = 6.45%), the compaction increased *RH* from 55% (value imposed when hydrating the mixture prior to compaction) to 62±2% after compaction. During infiltration, the value of *RH* at $h$ = 50 mm increased rapidly and reached a zero suction state after 4000 h. Afterwards, the signal of the *RH* probe at $h$ = 50 mm was lost, because resistive probes are known to fail at saturated relative humidity due to water vapour condensation. At $h$ = 250 mm, *RH* increased from 63 to 73% after 7500 h.

The wetter mixture used in Test T03 ($w_i$ = 9.95%) was prepared at *RH* = 90% and the initial *RH* measured in the column was 90±1%. The *RH* value at $h$ = 50 mm increased rapidly from 90 to 98% after 1000 h while *RH* at $h$ = 250 mm increased from 89 to 93% after 2400 h. As explained previously, the *RH* probe at $h$ = 50 mm failed after 1000 h once the humidity was saturated (*RH* = 100%).

The total suction (*s*) was calculated from *RH* using the Kelvin's law:

$$s = -(\rho_w RT/M_w) \ln(RH/100) \qquad [1]$$

where $R$ is the universal (molar) gas constant (8.31432 J/mol.K) ; $T$ is the absolute temperature (K); $M_w$ is the molecular mass of water vapour (18.016 g/mol) and $\rho_w$ is the water unit weight (1000 kg/m$^3$).

Figure 7 presents the suction isochrones for Test T01 with a time step of 200 h. At $t = 0$, the initial total suction of the compacted sample is approximately constant with a mean value of was 49.5±1.5 MPa. After starting the infiltration, the suction at the wetting face ($h = 0$) rapidly decreased to zero. Note that this zero suction is an imposed value and the real suction in the mixture in this level must took some time to decrease to zero. The total suction at $h = 50$ mm rapidly decreased to 4.1 MPa at $t = 2200$ h while the total suction at $h = 250$ mm remained high ($s = 45.5$ MPa at $t = 2200$ h).

The determination of the unsaturated hydraulic conductivity using the generalized Darcy's law according to the instantaneous profile method is detailed in Figure 8. In Figure 8*a*, the suction profiles at $t = 200$ and 600 h are plotted. The hydraulic gradient (*i*) is calculated as the slope of the isochrone (tangent of the suction profile at *h* and *t*) as follows:

$$i = ds/dh \qquad [2]$$

where *s* and *h* are expressed in m.

In Figure 8*a*, the hydraulic gradients at $h = 50$ mm, at $t = 200$ and 600 are respectively 25 000 and 53 400. The volume of water passing through the area located at $h = 50$ mm in the time period comprised between *t* and *t*+d*t* is computed by integrating the difference in the water content profiles at time t and *t*+d*t*.

In this work, the water content of the soil was calculated from the total suction based on the water retention curves obtained at constant-volume conditions on the same material by Loiseau et al. (2002). The following empirical equation was derived from Figure 2:

$$w = -5.9\log(s) + 17.7 \qquad [3]$$

Note that the total suction measured by *RH* probe is ranging from 4 to 100 MPa. The relationship between the logarithm of suction and water content obtained by Loiseau et al. (2002) can be correlated with a linear function in this range of suction. For this reason, a linear correlation was used to calculate the water content from the suction measured.

Using this correlation, the volumetric water content ($\theta$) can be calculated by:

$$\theta = w\rho_d/\rho_w \qquad [4]$$

In the infiltration test at constant-volume conditions, $\rho_d$ is assumed to be constant, $\rho_d = 2.0$ Mg/m$^3$. In Figure 8*b*, the volumetric water content profiles at $t = 200$ and 600 h are calculated from the suction profiles and plotted. The shaded area corresponds to the volume of water passing through the point $h = 50$ mm during the time period between 200 and 600 h.

The same procedures have been applied after various periods of time with a time increment of 100 h. The results in terms of water fluxes (*q*) and hydraulic gradients (*i*) are plotted in Figure 9 and Figure 10 respectively for distances from the wetted ends $h$ = 50, 100, 150, and 200 mm. The maximal measured flux $q$ is equal to 1.8 x $10^{-12}$ m$^3$/s. Calculations show that huge values of hydraulic gradients are mobilised, with values of $i$ at $h$ = 50 mm increasing rapidly from 0 to 54 000 in the first 550 hours and subsequently decreasing to 19 000 at $t$ = 2250 h. These huge values of hydraulic gradients are to be related to the significant changes in suction with respect to the distance to the wetted end that occur during hydration.

In Figure 11, the calculated flux $q$ is plotted versus the hydraulic gradient $i$ at distances $h$ = 50, 100, 150, and 200 mm from the wetted end, according to the results presented in Figure **9**. It can be observed that various $q$-$i$ relationships are observed depending on the position considered, with not simply linear shapes like in cases where Darcy's law is respected (the hydraulic conductivity $k$ being given by $k = (q/i)/A$ where $A$ is the cross-sectional area equal here to 0.00196 m$^2$).

In Figure 12, the water flux $q$ is plotted versus the hydraulic gradient $i$ for each value of suction (according to the results presented in Figure 11). It can be observed that, for a given suction, the $q$-$i$ relationship is bilinear with two slopes; the slope at high gradients being larger than that at low gradients. This bilinear relationship of $q$-$i$ has often been observed when measuring the saturated hydraulic conductivity of clayey soils (Dixon et al. 1992).

In Figure 13, the values of unsaturated hydraulic permeability $k$ is plotted versus suction according to calculations made at various distances from the wetting point with $h$ = 50, 100, 150 and 200 mm. Different curves are obtained in different levels. This is the consequence of the bilinear relationship of $q$-$i$ observed above (Figure 12), evidencing the effect of water gradient. To analyse the water flow using Darcy's law, one of the possibilities is to use the term "critical gradient", i.e., to consider water flow only when the linear segment at higher water flux (Figure 12) is reached (more details about this method are presented in Dixon et al. 1992). This analysis using critical gradient leads to a unique relationship, independent of the measurement level. This relationship refers to T01 in Figure 13. It can be observed that $k$ decreases from 12 x $10^{-14}$ to 3 x $10^{-14}$ when suction decreases from 50 to 25 MPa. On the contrary, $k$ slightly increases for further decreases in suction from 25 to 15 MPa,.
Unlike in low plasticity unsaturated soils in which no significant strain is observed during infiltration and in which the coefficient of permeability is increasing during infiltration, it can be observed that the $k$-$s$ relationship is not unique in swelling soils with constant-volume conditions. This non unicity obviously add a degree of complexity when trying to model and calculate water transfers in this material.

Similar procedures were applied to determine the unsaturated hydraulic conductivity from tests T02 and T03 and all results are presented together in Figure 21.

The results obtained from the infiltration test T04 in which swelling was allowed are presented in Figure 14 and Figure 15. In Figure 14, the values of $RH$ measured by each probe are plotted versus time $t$. The test was conducted during 1600 h with four $RH$ probes embedded at distances $h_i$ = 10 (S1), 40 (S2), 65 (S3) and 95 mm (S4) from the wetted end. In general, the shape of the curves is comparable with that of the constant volume infiltration tests (see Figure 6). The *RH* measured by probe S1 located closest to the wetting face ($h_i$ = 10 mm) increased quickly and reached a zero suction condition after only 50 h (the probe afterwards failed due to vapour condensation). Probe S2 failed at 800 h and the Probe S3 failed at 1000 h.

In Figure 15, the evolution of the axial and radial strain that were monitored by the displacement gages during infiltration are plotted at various times. After 1400 h, the radius at 10 mm from the wetted end increased from 25 to 30 mm (corresponding to a radial swelling strain $\varepsilon_{rs}$ = 20%) and the distance between S1 and the wetting face (*h*) increased from 10 to 18 mm (80%). The total length of the soil specimen (distance between the top and the wetting face) increased from 100 to 118 mm (18%).

The results presented in Figure 15 enable calculating the total volume of the soil specimen at various times. In Figure 16, the volume change of the soil specimen, the total volume of water inlet, and the total volume of air outlet are plotted together versus time. It appears that after starting the infiltration, water was absorbed in the soil at a rate of 0.1 cm$^3$/h. After 800 h (33 days), the soil stopped to adsorb water and the volume of the soil stabilized. The air volume measured at the air outlet increased and stabilized at 400 h with a total value of 4.5 cm$^3$.

The determination of the coefficient of unsaturated hydraulic conductivity from data of Test T04 is similar to that of T01 presented previously. The results presented in Figure 14 were first used to plot the suction profiles at various times. According to the water retention curve obtained by Loiseau et al. (2002) during wetting under free-swell condition (Figure 2), the following correlation was adopted to calculate the water content *w* from suction *s*:

$$w = 24.6 \, s^{-0.32} \qquad [5]$$

with *s* in MPa. This correlation as well as the results on volume change presented in Figure 15 were then used to plot the volumetric water content profiles at various times.

In order to check the validity of these analyses, the total volume of water passing through the area where the probe S1 is located was calculated from the volumetric water content profiles. This volume is plotted as *V1* versus time in Figure 17. *V2* corresponds to the increase of volume of water in the soil between the wetting face and the location of S1. The sum of these volumes $V_{calcul}$ = V1 + V2 corresponds to the estimated volume of water absorbed by the soil specimen during infiltration. In Figure 17, the measured volume of water inlet, $V_{measured}$, is also plotted versus time. It can be observed that $V_{calcul}$ is in good agreement with $V_{measured}$, confirming the validity of the analyses performed on Test T04.

Using the suction and volumetric water content profiles, the hydraulic conductivity at each location of probe (*h* = 10, 40, 65 and 95 mm, Figure 18) can be determined by applying the instantaneous profile method. As seen in the Figure 18, the *k-s* plots determined at the different locations are different. In order to estimate the suction effect on the changes of *k* in free-swell condition, the following correlation was proposed according the results presented in Figure 18:

$$k = 3 \times 10^{-13} s^{-0.8} \quad [6]$$

That enables estimating the changes of $k$ when $s$ is decreasing:

$$dk/ds = 2.55 \times 10^{-13} s^{-1.85} \quad [7]$$

For further analyses, $k$ is plotted versus the void ratio, $e$, in Figure 19. In can be observed that $k$ increases with increasing void ratio. The following correlation can be proposed:

$$k = 9 \times 10^{-16} e^{9.4e} \quad [8]$$

Equation [8] was used to estimate the hydraulic conductivity at $e$ = 0.2, 0.3, 0.4, and 0.5 (at initial suction equal to 37 MPa). Equation [7] was then applied to calculate the changes in $k$ when suction is decreasing for each value of $e$. All the results are plotted in Figure 20. These results show the increase of $k$ when $e$ is increasing or $s$ is decreasing.

In Figure 21, $k$ corresponding to $e$ = 0.34 calculated from the test T04 is plotted together with the results obtained from tests T01, T02, and T03 on the soil samples with the same void ratio $e$ = 0.34. Loiseau et al. (2002) measured the $k_{sat}$ of this material at the same void ratio ($e$ = 0.34) by injecting water at constant pressure and monitoring the rate of inlet water flow. During the pressure increase, when the hydraulic gradient increased from 0 to 25000, a decreasing water flux with water gradient was observed; under higher gradients (up to 31000), a linear q – i relation was observed. The same linear relation was observed when the hydraulic gradient was decreased to 7500. The initial hydraulic conductivity ($k_{ini}$ = 4 x 10$^{-13}$ m/s) is thus higher than the final one ($k_{final}$ = 6 x 10$^{-14}$ m/s). These values are equally brought in Figure 21.

## *Discussion*

It has been observed that the initial *RH* values measured in the infiltration tests are slightly higher than the *RH* applied to wet the sand/bentonite mixture prior to compaction. That means the compaction increased *RH* in the soil (or decreased the total soil suction). Blatz and Graham (2000) noted equally such reduction of total suction in compacted sand/bentonite mixture when the net mean stress is increasing. Yahia-Aissa et al (2001) indicated that there are in general two main parts of suction in expansive soils: one related to the physico-chemical interaction between water and clays particles and another related to the capillarity effect. It is believed that compaction can only affect the capillarity part and, therefore, the capillary suction in the studied mixture is not negligible compared to the physico-chemical one.

The changes of *RH* with time at various locations (Figure 3 and Figure 14) are similar with the observations made by Lemaire et al. (2004) and Kröhn (2003*a*). In Test T01, *RH* at the upper end ($h$ = 250 mm) started to increase at 1000 h while *RH* at $h$ = 50 mm remained at 90% (corresponding to $s$ = 12 MPa, $w$ = 11.3% and $S_r$ = 90%). For others tests, *RH* at the points located far from the wetting face started to increase quickly after starting the infiltration. Lemaire et al. (2004) monitored the water content at different locations on the soil specimen during infiltration and observed the same phenomenon. Infiltration test performed by Kröhn (2004*a*) on compacted air-dry bentonite ($\rho_d$ = 1.5 Mg/m$^3$) showed that the water content located at 10 cm from the wetting face started to increase only after 11 days.

Kröhn (2004*a*) and Börgesson et al. (2001) determined the water content profiles during infiltration test by cutting the soil specimen in slices. The number of profiles obtained was then limited. With the method applied in the present work as well as in the work of Lemaire et al. (2004) where suction (or water content) was monitored in a continuous fashion, the suction profile (or water content profile) can be established for any time during the test. That enables improving the accuracy when determining the hydraulic conductivity. The disadvantage of the continuous measurement is that the number of measurements on each profile is limited by the number of probes.

In the present work, the instantaneous profile method has been applied to determine the unsaturated hydraulic conductivity. After Daniel (1982) this method does not allow determining the portion of water flow in liquid phase and in vapour phase. In addition, as the total suction was measured, the osmotic suction gradient and matric suction gradient were not distinguished.

According to Darcy's law, a linear relationship between water flow rate ($q$) and hydraulic gradient ($i$) must be observed when undertaking tests for hydraulic conductivity measurement. After Dixon et al. (1999), the Darcy's law is commonly accepted in coarse-grained natural soils but it should be applied with care on low-permeability clay. Zou (1996) explained this non-linear relationship by the activation energy of pore liquid that is much higher in fine-grained soils than in coarse-grained soils. This explanation can be also applied to the different *k-s* relationships obtained from different probes (Figure 13) and determined directly from the results presented in Figure 11. This activation energy can be related to the notion of critical gradient. Using *q-i* relationships plotted for each suction (Figure 12), this critical gradient can be determined, allowing the correction of the hydraulic conductivity determined (see Dixon et al. 1992 for more details about the correction). The corrected *k-s* curve is presented in Figure 13 and it can be observed that it situates above the non corrected curves.

The soil specimens used in Tests T01 and T04 are similar ($\rho_d$ = 2.0 Mg/m$^3$; $w_i$ = 7.7% for T01 and 8.2% for T04). The initial water content of T04 was slightly higher than that of T01, that explains why the initial *RH* measured in T04 (74±2%, Figure 14) was higher than in T01 (70±1%, Figure 6). However, the infiltration rate in Test T04 is much higher than in Test T01; *RH* at $h$ = 100 mm reached 90% after 2400 in Test T01 (Figure 6) while *RH* at $h_i$ = 100 mm (Probe S4) reached 90% after 1100 h in Test T04 (Figure 14). Yong and Mohamed (1992) performed infiltration tests on compacted expansive soil and also observed that the wetting-front in the case of axial free-swell conditions advanced more quickly than in the test at constant-volume conditions. This clearly shows that the soil microstructure changes are different at the two conditions.

The microstructure of the studied mixture has been investigated by Cui et al. (2002) using mercury intrusion porosimetry and scanning electron microscopy. Compacted soil specimens were wetted under constant-volume conditions forward different suctions ranging from 57 MPa to 0. The pore size distribution curve at $s$ = 57 MPa showed a bimodal shape with the macro-pore family at 0.35 μm and micro-pore family at 0.025 μm. Wetting under constant-volume conditions decreased both the volume and the size of macro-pores; at $s$ = 0, the macro-pore family disappeared. These observations can be used to explain the different infiltration rates between constant-volume and free-swell conditions. The clay aggregates swell during wetting, thus reducing the volume of macro-pore family under constant-volume conditions. On the contrary, in free-swell conditions, swelling of aggregates may increase the volume of macro-pore family, thus increasing the water flow rate or hydraulic conductivity.

In Figure 18, the *k-s* relationships were determined for each probe using Darcy's law without the correction, including therefore the gradient effect. In fact, during wetting under free-swell conditions, *k* changed with changes in suction, hydraulic gradient and void ratio, hence plotting *q-i* curve for each suction (Figure 12) does not help correct the effect of gradient. However, the proposed correlations (best fit) can give estimation about the *k-s* relationship (Figure 18) and the *k-e* relationship (Figure 19). The *k-s* curves plotted in Figure 20 according to these two correlations would correspond to a soil having unchanged microstructure during wetting.

In Figure 21, it can be observed that $k_{sat}$ determined by Loiseau et al. (2002) reduces with time. The $k_{sat}$ decrease during the tests was equally observed by Haug and Wong (1992) and Hoffmann et al. (2007) on compacted expansive soils. Lloret and Villar (2007) mentioned that the permeability of compacted bentonite measured under water saturated conditions is much lower than that measured by gas under unsaturated or dry conditions. The reduction of the volume of macro-pore family during wetting observed by Cui et al. (2002) can explain this phenomenon of hydraulic conductivity decrease.

From the analysis above it can be concluded that $k_{ini}$ corresponds to the initial microstructure of the compacted sand/bentonite mixture. Note that the plot of test T04 in Figure 21 corresponds to the *k-s* curve of a soil sample having a microstructure similar to the initial microstructure of a compacted soil sample at $e$ = 0.34. Moreover, the soils tested by Loiseau et al. (2002) and in Test T04 have similar initial state ($e$ = 0.34, $w_i$ = 8.0±0.3%). In Figure 21, the results of Test T04 are plotted for suctions higher than 4 MPa because the *RH* failed at lower suctions. It can be observed that the *k-s* curve determined from Test T04 trends to reach $k_{ini}$ when suction reduces to zero. This confirms that the microstructure of the soil tested in the two tests was similar, leading to a unique $k_{ini}$ value.

On the other hand, $k_{final}$ corresponds to the final microstructure of the soil after wetting under constant-volume conditions; the macro-pore family was almost eliminated. Interestingly, the *k-s* curves obtained from the three tests performed at constant-volume conditions (T01, T02 and T03) trend to join $k_{final}$ when suction reduces to zero, whatever the water content.

Upon wetting, the suction is decreasing and water conductivity $k_{unsat}$ is in general increasing because water retention force is reduced (Daniel 1982; Chiu and Shackelford 1998; Benson and Gribb 1999). On the other hand, wetting under constant-volume conditions reduces the volume of the marco-pore family (Cui et al. 2002), giving rise to a decrease of $k_{unsat}$. These two opposing mechanisms explain the results on constant-volume conditions presented in Figure 21: during wetting, *k* reduced initially because of the reduction of volume of macro-pore; in Test T01 and T02, *k* started to increase with the decrease of suction from *s* = 22 MPa.

## *Conclusion*

This work aimed at determining the unsaturated hydraulic conductivity of a compacted sand/bentonite mixture. Three infiltration tests were undertaken under either constant-volume conditions or free-swell conditions. The total suction changes were monitored at different locations along the cylindrical soil specimen and the instantaneous profile method was applied to calculate the hydraulic conductivity.

The tests under constant-volume conditions were performed at three values of moulding water content. When suction was decreasing (during infiltration), the hydraulic conductivity presented an initial decrease followed by an increase after a certain suction. This phenomenon can be explained by the fact that wetting under constant-volume condition reduced the volume of macro-pore family in the soil, thus decreasing the soil permeability. When the macropores disappeared, the suction became the only factor which drove the water flow and the hydraulic conductivity increases with further decrease in suction.

During the infiltration test under free-swell conditions, in addition to the measurement of the total suction, the volume changes of the soil specimen were equally monitored by displacement transducers. The experimental results enabled calculating the unsaturated hydraulic conductivity for a given suction and void ratio. This hydraulic conductivity corresponds to the equivalent value of a soil having a microstructure unchanged during wetting.

Loiseau et al. (2002) determined the saturated hydraulic conductivity of the same material under constant-volume conditions and showed that water conductivity was decreasing during the test. This can be explained by the decrease in volume of the macro-pore family due to wetting under constant-volume conditions. It was observed that the saturated hydraulic conductivity determined in the beginning of the test, when microstructure was still not changed, is in good agreement with the results obtained from the infiltration test under free-swell condition. Interestingly, the saturated hydraulic conductivity determined at the end of the test, when the macro-pore family had been already eliminated, is in good agreement with that obtained from the infiltration tests under constant-volume conditions. These observations clearly show the primary role played by the soil microstructure in the water flow within the soil.

## *Acknowledgements*

The first author is grateful to the National Natural Science Foundation of China for its supports (No. 40728003).

**Table 1. Geotechnical parameters of Kunigel-V1 clay. (After Komine 2004)**

| Type | Sodium bentonite |
|---|---|
| **Particle density (Mg/m$^3$)** | 2.79 |
| **Liquid limit (%)** | 474 |
| **Plastic limit (%)** | 27 |
| **Activity** | 6.93 |
| **Clay (< 2μm) content (%)** | 64.5 |
| **Montmorillonite content (%)** | 48 |
| **Cation exchange capacity (meq./g)** | 0.732 |
| **Exchange capacity of Na$^+$ (meq./g)** | 0.405 |
| **Exchange capacity of Ca$^{2+}$ (meq./g)** | 0.287 |
| **Exchange capacity of K$^+$ (meq./g)** | 0.009 |
| **Exchange capacity of Mg$^{2+}$ (meq./g)** | 0.030 |

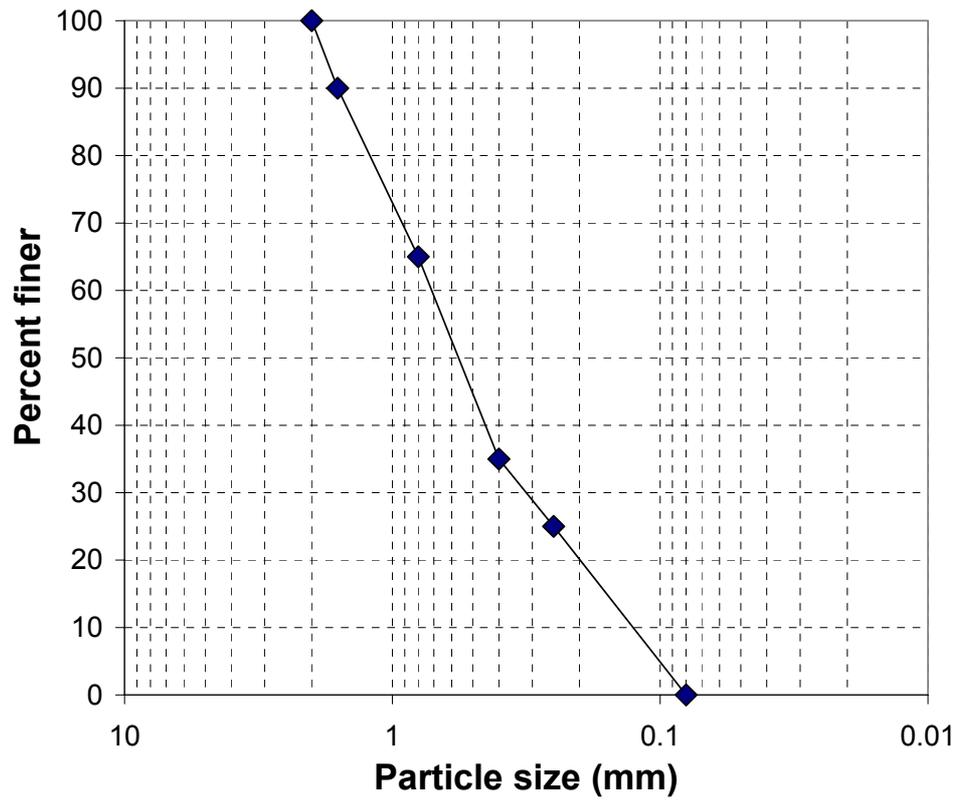

**Figure 1. Particle-size distributions for sand.**

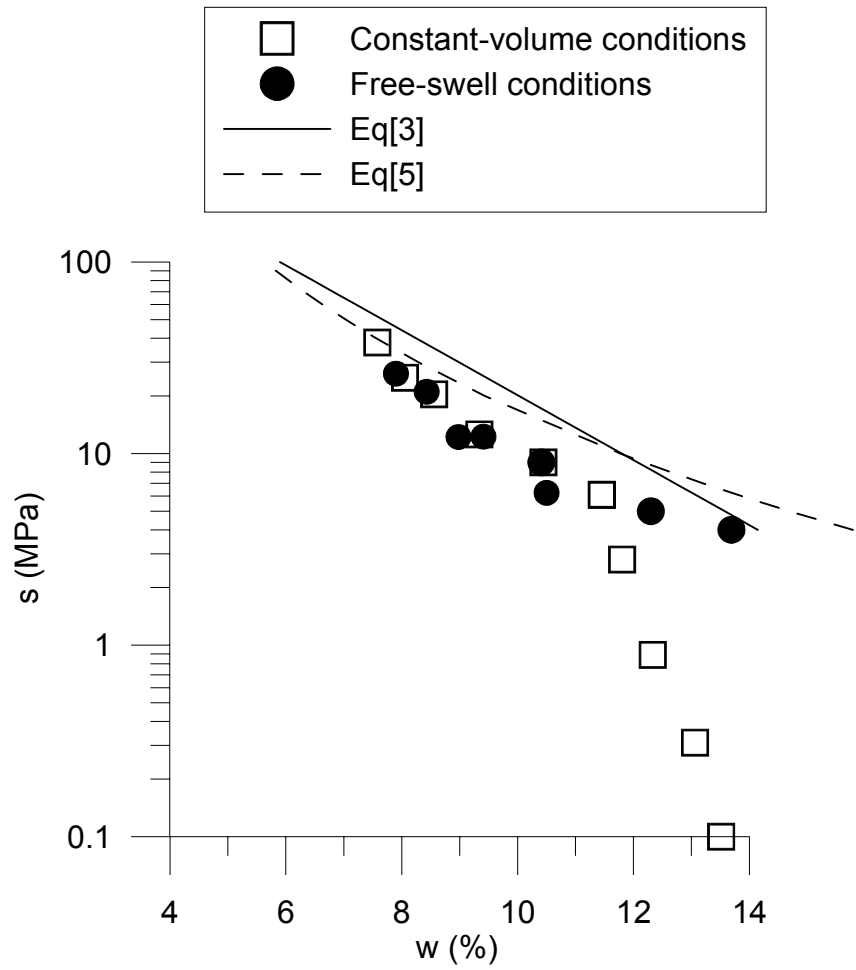

**Figure 2. Water retention curves in constant-volume and free-swell conditions (After Loiseau et al. 2002).**

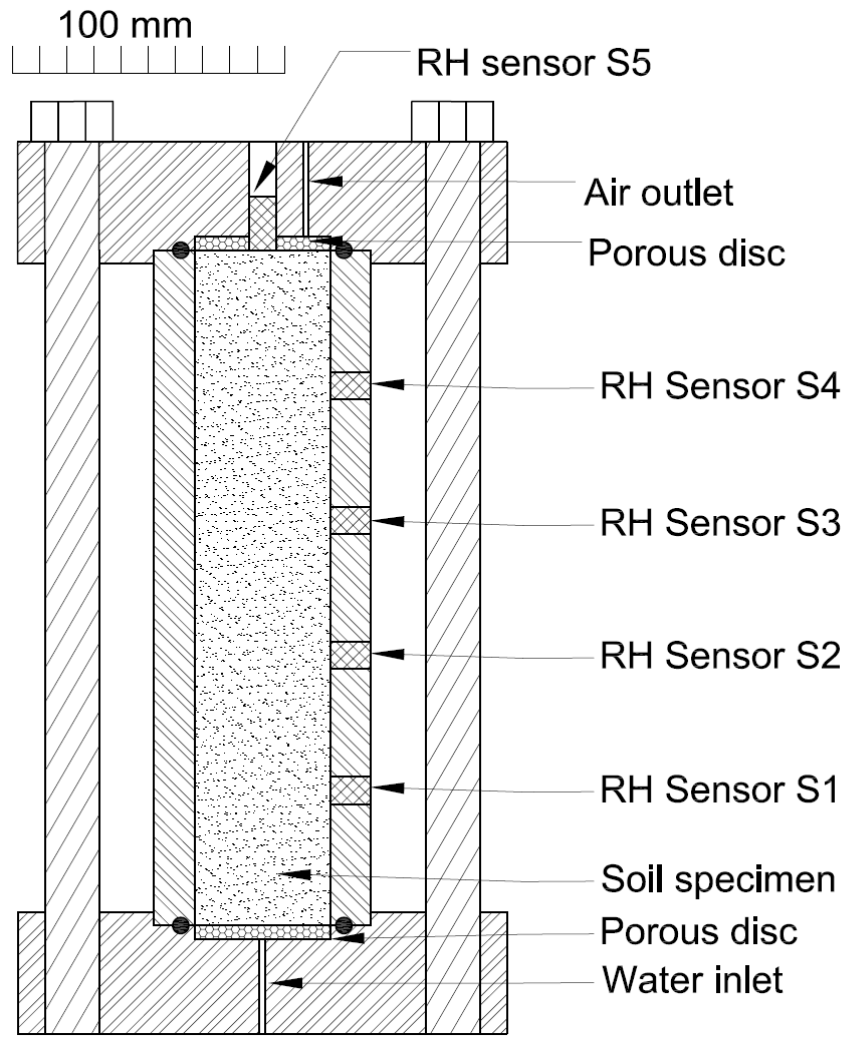

**Figure 3.** Schematic diagrams of the infiltration test with no volume changes.

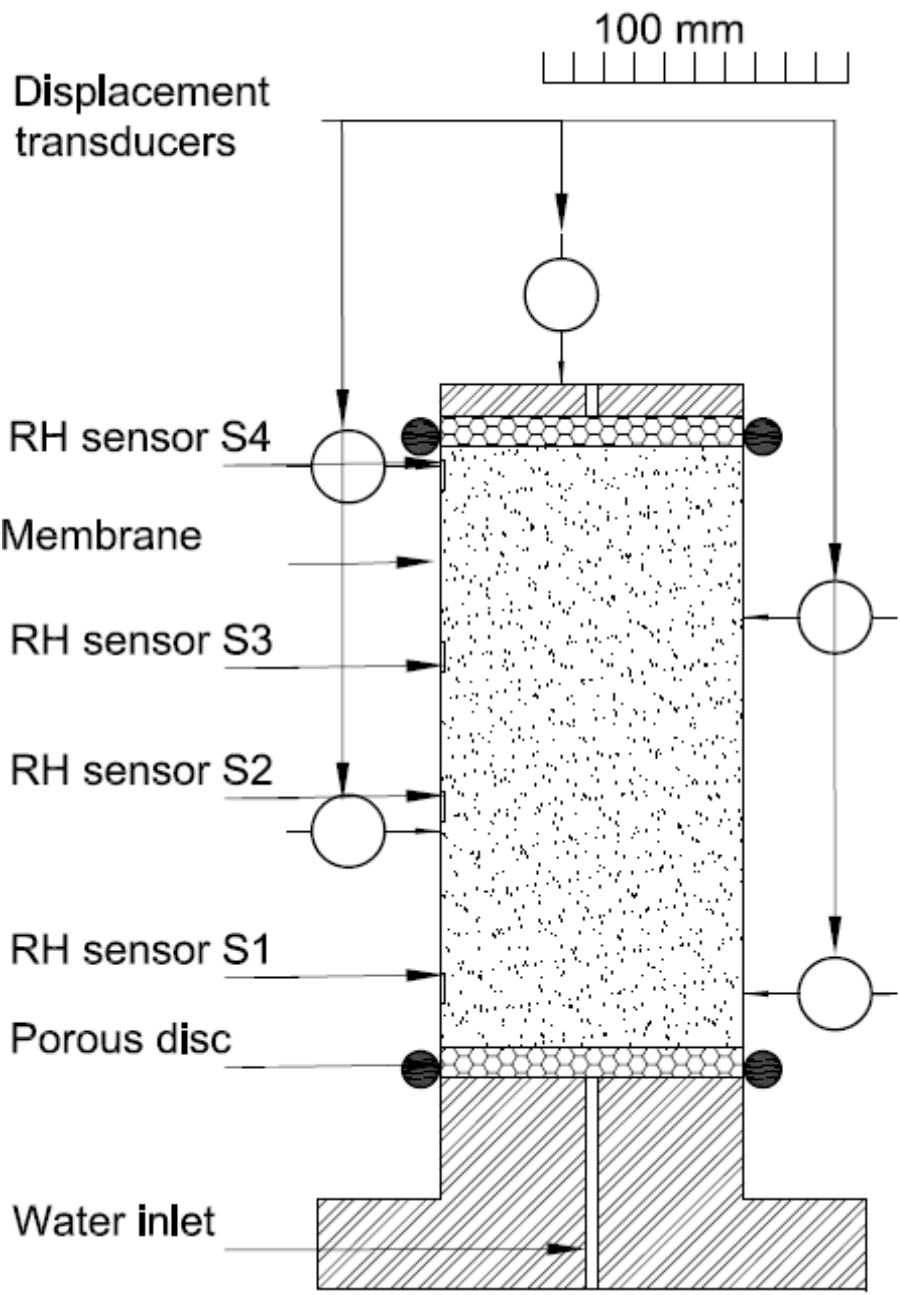

**Figure 4. Schematic diagrams of the infiltration test with swelling allowed.**

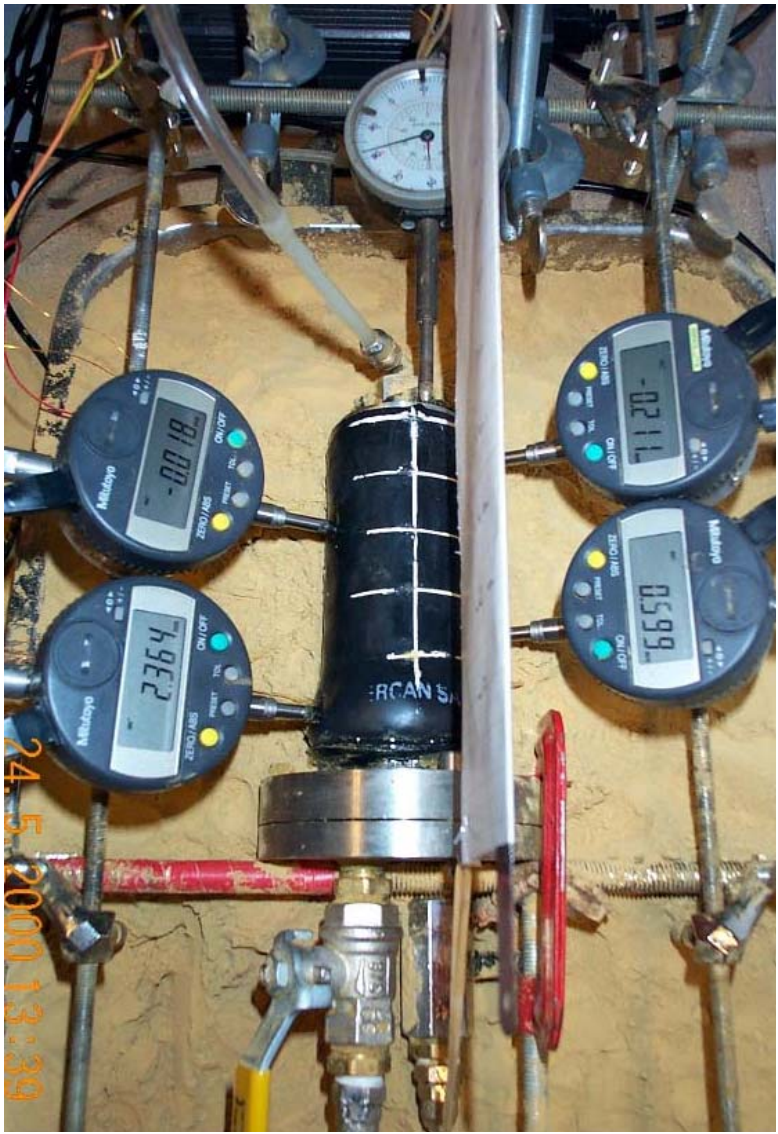

**Figure 5. Picture of the infiltration test with swelling allowed.**

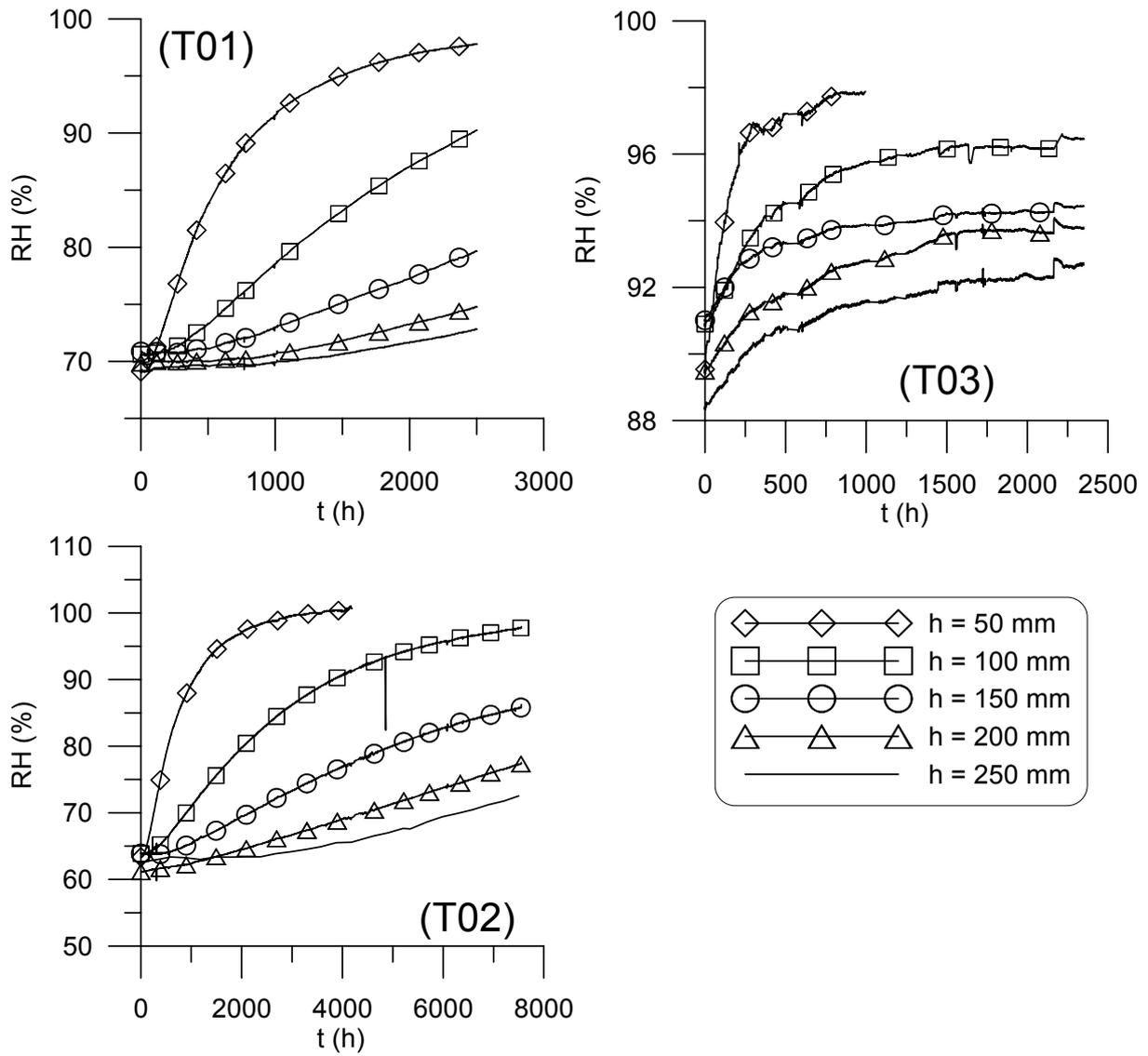

**Figure 6. Relative humidity changes versus time for infiltration tests under constant-volume condition. Tests T01, T02, and T03.**

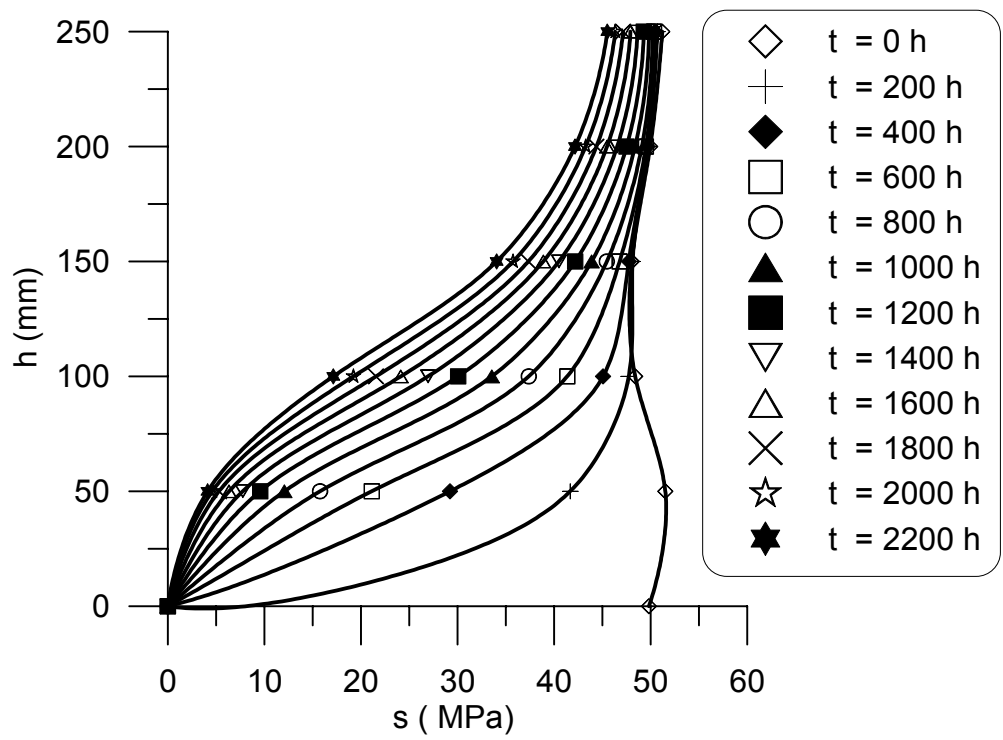

**Figure 7. Test T01. Suction profile at different times.**

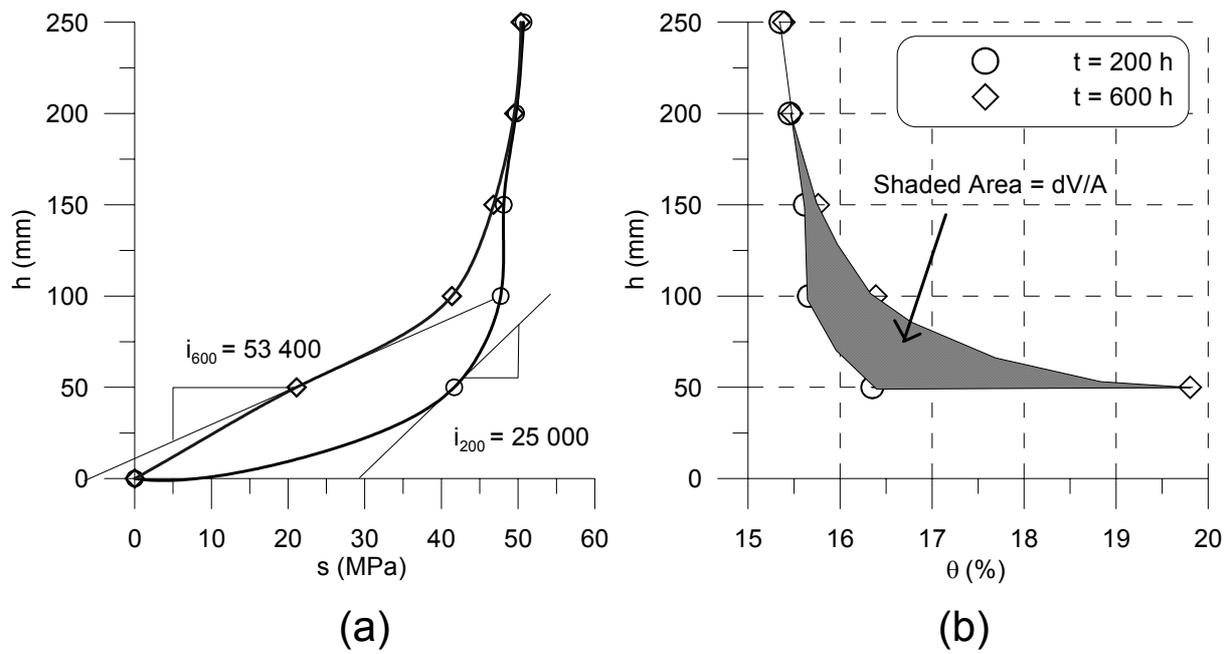

**Figure 8. Test T01. Determination of unsaturated hydraulic conductivity: (a) hydraulic gradient; (b) rate of water flow.**

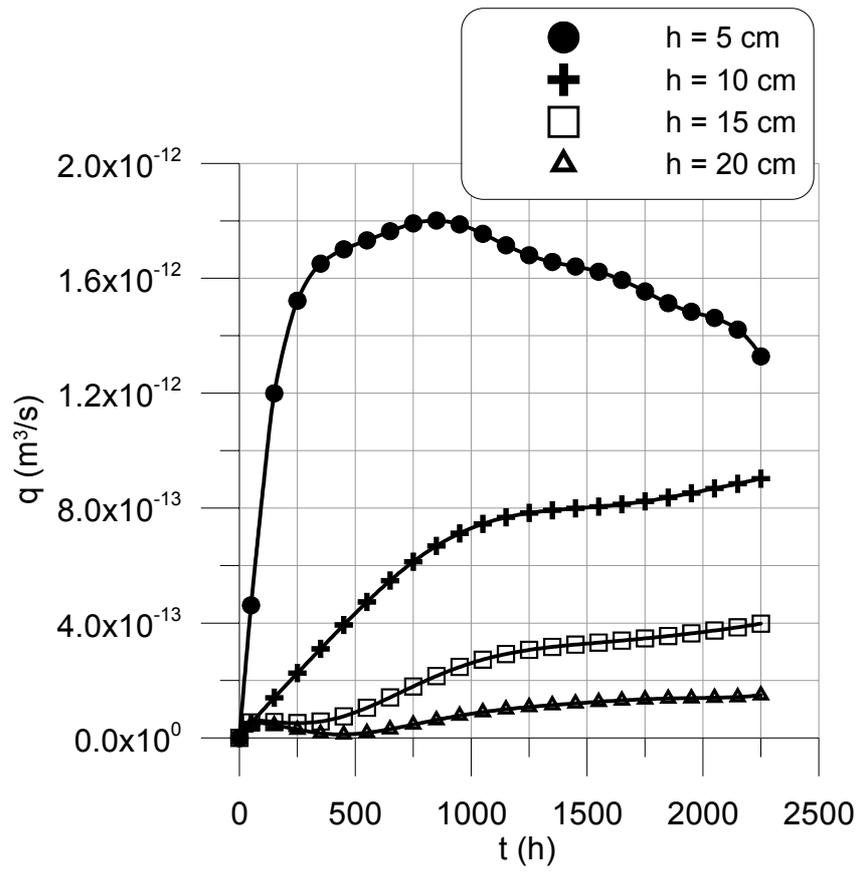

**Figure 9. Test T01: Evolution of water flux during infiltration.**

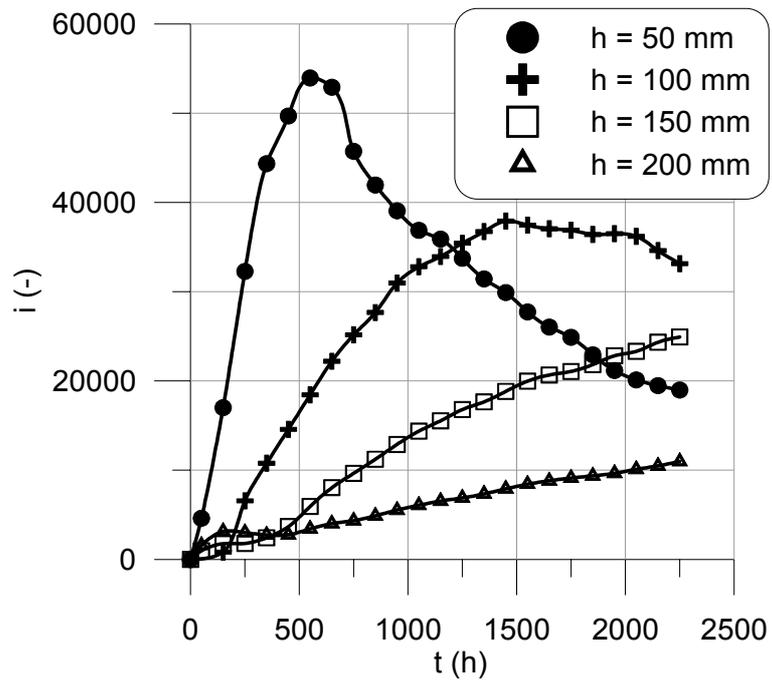

**Figure 10. Test T01. Evolution of hydraulic gradient during infiltration.**

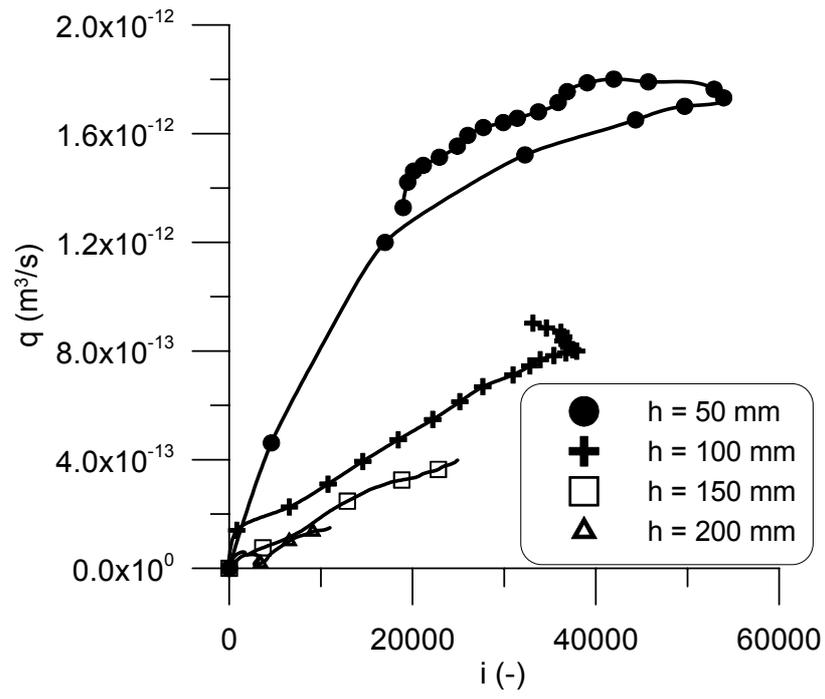

**Figure 11. Test T01. Water fluxes versus hydraulic gradient for each location.**

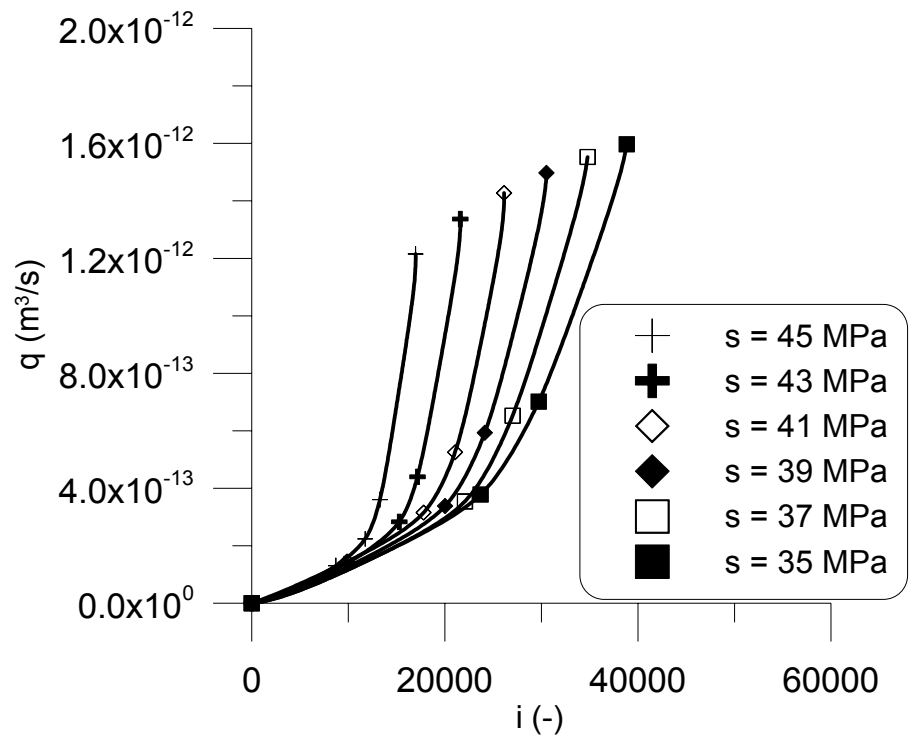

**Figure 12. Test T01. Water fluxes versus hydraulic gradient for each suction value.**

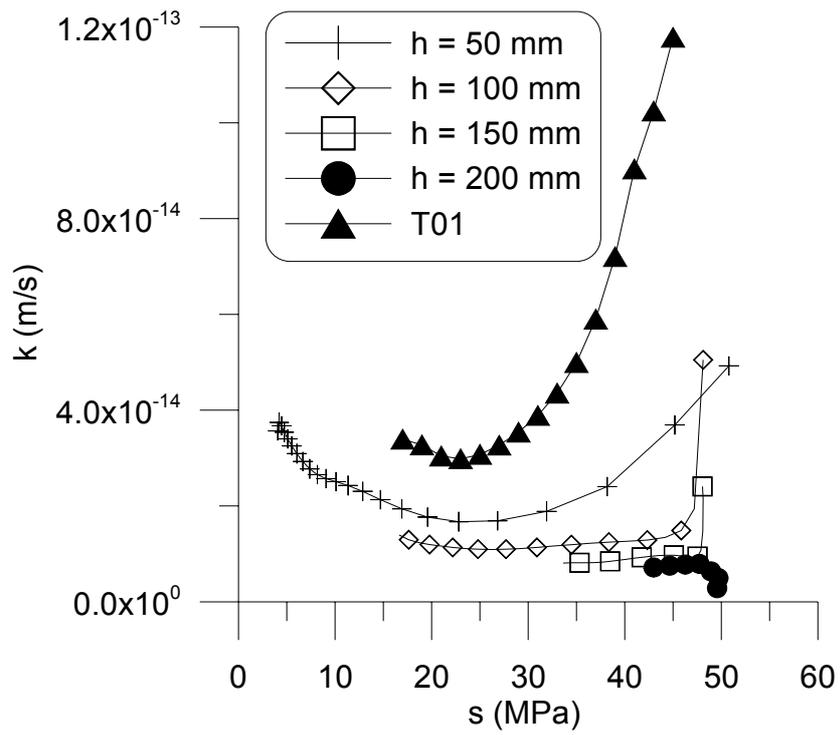

**Figure 13. Test T01. Hydraulic conductivity versus suction.**

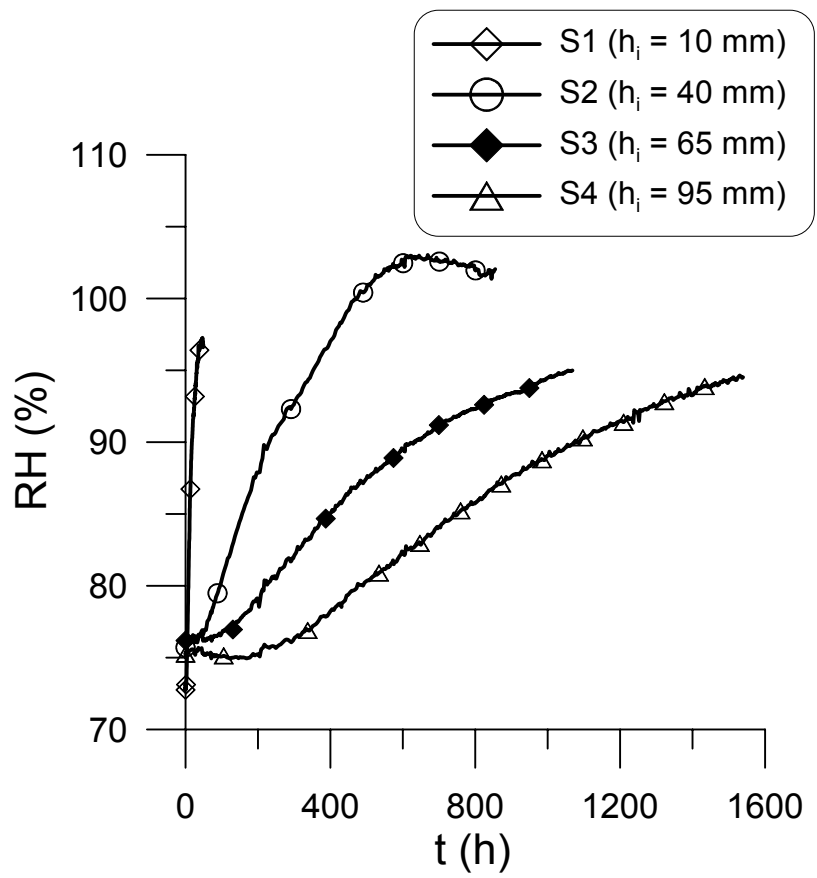

**Figure 14. Test T04. Relative humidity versus time at different locations.**

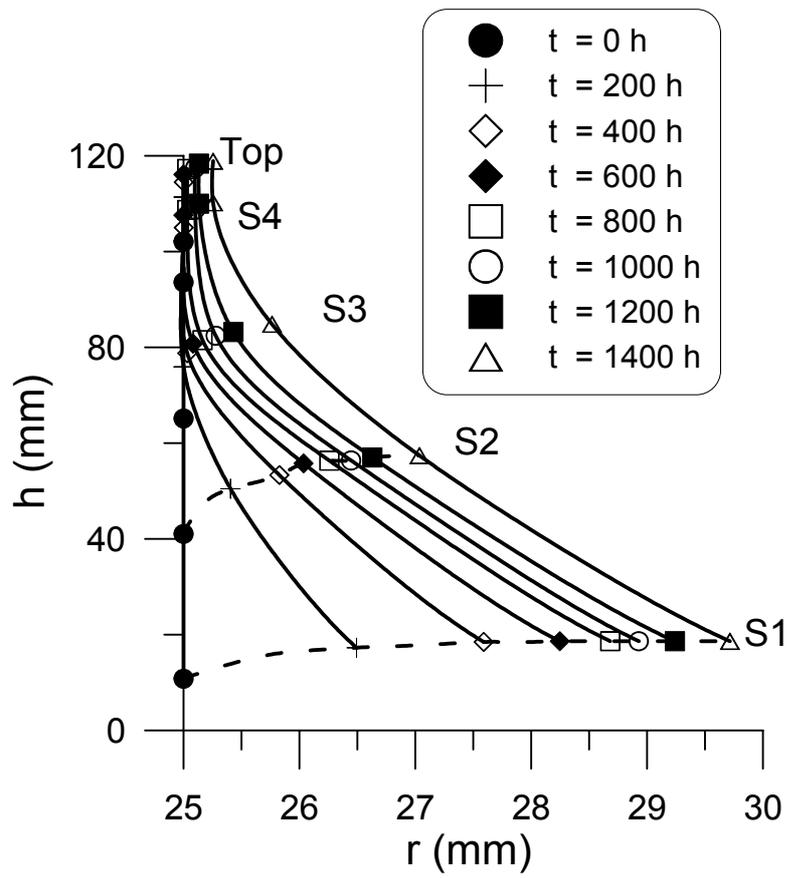

**Figure 15. Test T04. Axial and radial deformation at various times.**

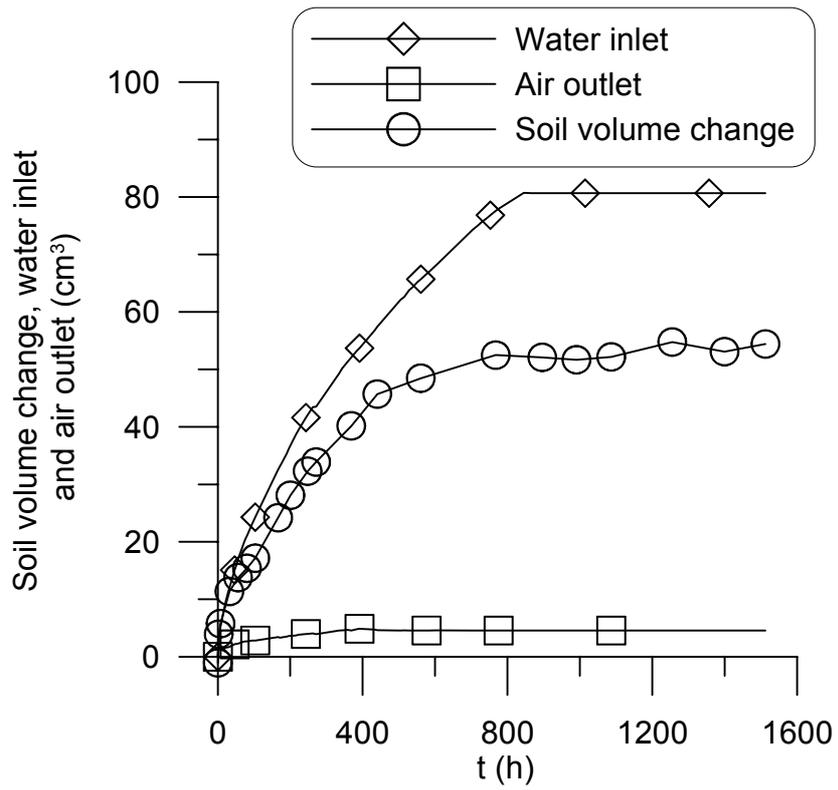

**Figure 16. Test T04. Soil volume change, water inlet and air outlet versus time.**

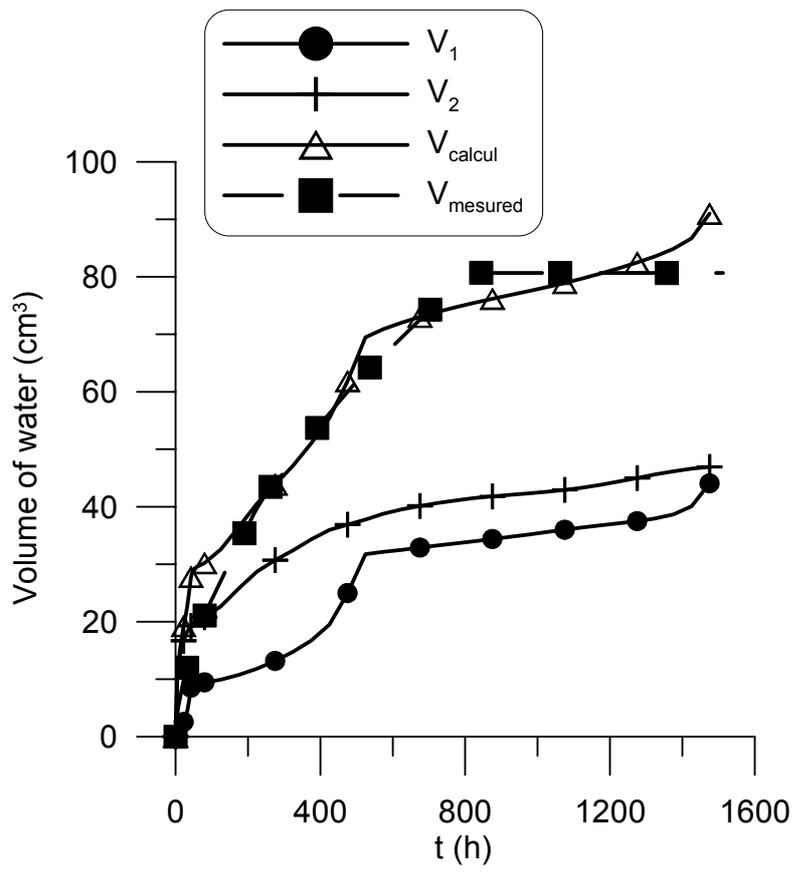

**Figure 17. Test T04. Volume of water inlet estimated and measured.**

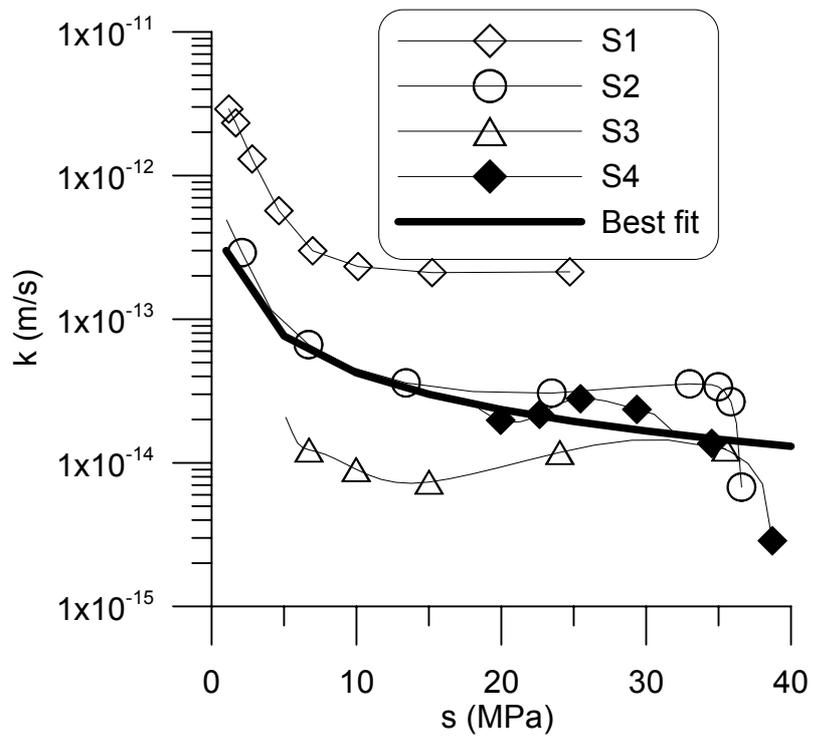

**Figure 18. Test T04. Hydraulic conductivity at various locations of RH sensors versus suction.**

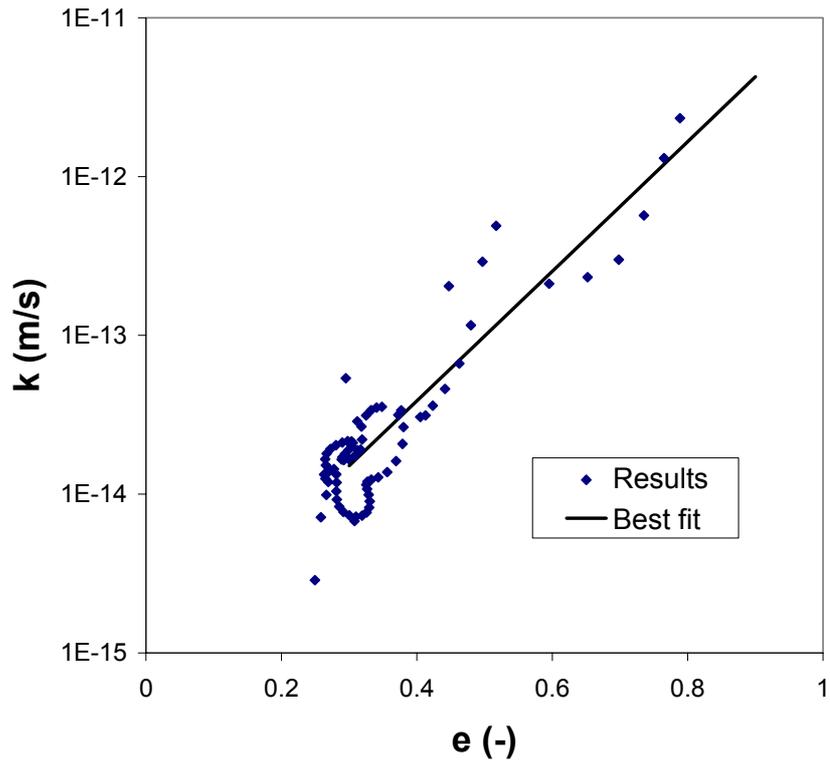

**Figure 19. Test T04. Hydraulic conductivity versus void ratio.**

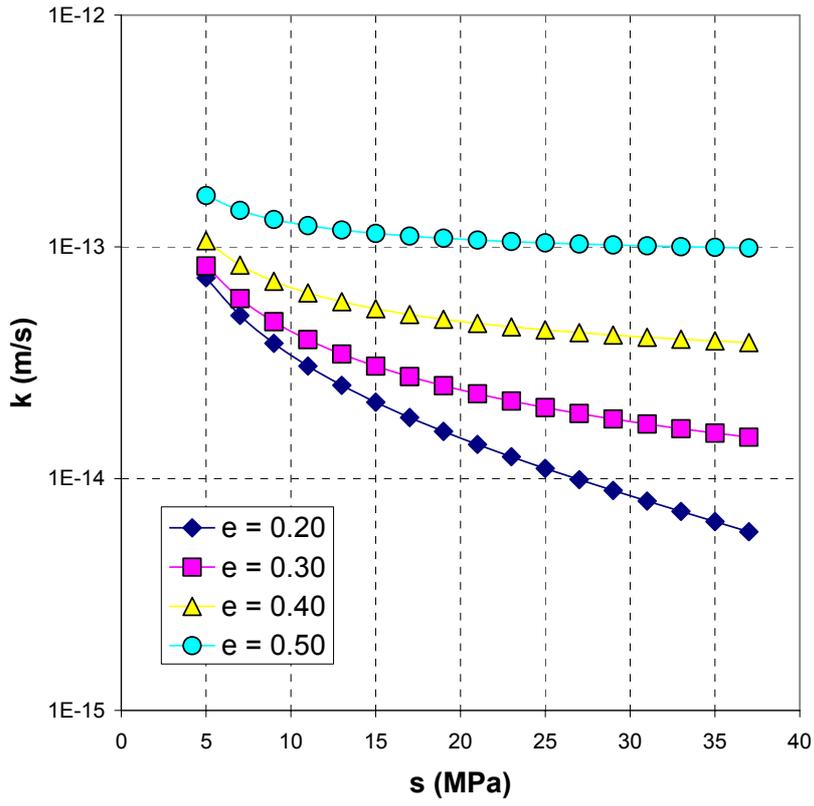

**Figure 20. Test T04. Equivalent hydraulic conductivity versus suction for various void ratio values.**

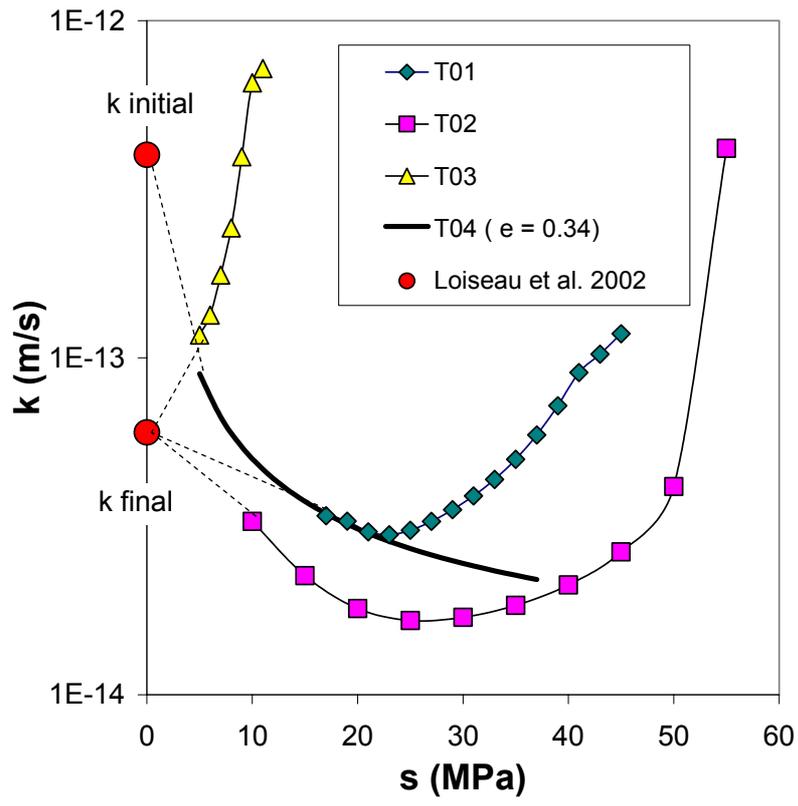

**Figure 21. Hydraulic conductivity versus suction for all the tests.**